\documentclass{article}
\usepackage{report,times}
\usepackage{graphicx}
\usepackage{subcaption}
\usepackage{tabularx}
\usepackage{booktabs}
\usepackage{layouts}
\usepackage{nicefrac}
\usepackage{hyperref}
\usepackage{url}

\def\sS{\mathcal{S}}

\title{A Measure of Research Taste}

\author{Vladlen Koltun \& David Hafner\\
Intelligent Systems Lab, Intel
}

\iclrfinalcopy
\begin{document}

\maketitle

\begin{abstract}
Researchers are often evaluated by citation-based metrics. Such metrics can inform hiring, promotion, and funding decisions. Concerns have been expressed that popular citation-based metrics incentivize researchers to maximize the production of publications. Such incentives may not be optimal for scientific progress. Here we present a citation-based measure that rewards both productivity and taste: the researcher's ability to focus on impactful contributions. The presented measure, CAP, balances the impact of publications and their quantity, thus incentivizing researchers to consider whether a publication is a useful addition to the literature. CAP is simple, interpretable, and parameter-free. We analyze the characteristics of CAP for highly-cited researchers in biology, computer science, economics, and physics, using a corpus of millions of publications and hundreds of millions of citations with yearly temporal granularity. CAP produces qualitatively plausible outcomes and has a number of advantages over prior metrics. Results can be explored at \url{https://cap-measure.org/}
\end{abstract}

\section{Introduction}

Are we publishing too much? The annual production of scientific articles is increasing exponentially~\citep{Dong2017,Fortunato2018}. While science is thriving, the increasing rate of publication has been a cause for concern. \citet{Margolis1967} wrote ``The present scientific explosion gave rise to more than a proportional publication explosion, which not only reflects the scientific explosion but has its own dynamics and vicious circles. [...] as with any other type of inflation, the problem worsens: the more papers are written, the less they count for and the greater is the pressure to publish more.''. This ``publication inflation''~\citep{London1968} led to the introduction of ironic concepts such as ``Least Publishable Unit''~\citep{Broad1981} and the ``publon''~\citep{Feibelman2011}. \citet{Altman1994} calls for ``less research, better research, and research done for the right reasons''. \citet{GemanGeman2016} lament that ``Academia has largely become a small-idea factory. Rewarded for publishing more frequently, we search for `minimum publishable units.'". \citet{Frith2020} writes ``When I was first hired, I had proudly published one paper, and ten years later I had accumulated a mixed bunch of 28 publications, none highly cited. I would have no chance today.''. \citet{Sarewitz2016Nature} suggests that ``We can start by publishing less, and less often''.

One factor that may be affecting publication rates in the last two decades is the broad adoption of citation-based metrics that measure a researcher's output, such as the $h$-index~\citep{Hirsch2005}. The $h$-index is used by committees that decide who is hired, promoted, and funded~\citep{Abbott2010,McNutt2014,Hicks2015}. People are known to alter their behavior to optimize performance measures even to the detriment of the organization's mission~\citep{Ridgway1956,Campbell1979}. Since the $h$-index has been institutionalized as a measure of performance, concerns have been raised that such metrics alter the behavior of scientists in ways that do not serve scientific progress~\citep{Lawrence2007,Lawrence2008,Abbott2010,GemanGeman2016,EdwardsRoy2017}.

One hypothetical solution is to abandon quantitative measurement of scientists' work~\citep{Lawrence2008,GemanGeman2016}. Perhaps whenever a funding decision needs to be made, the committee can read a small number of publications selected by the scientist and assess the quality of the work based on their expertise in the field~\citep{Lawrence2008,Hicks2015,GemanGeman2016}. We agree that in-depth assessment based on the content of the work is highly desirable, but anticipate that bibliometric indices will continue to be widely used. Calls for their abandonment have been made, yet their use is as widespread as ever, with the $h$-index prominently displayed on researcher profiles in scientific discovery platforms such as Google Scholar~\citep{HarzingAlakangas2016}, Semantic Scholar~\citep{Ammar2018}, and AMiner~\citep{Wan2019}. Committees that make decisions that can affect a scientist's career do not always have sufficient time or domain expertise to carefully read the scientist's papers and the related literature. When a funding committee needs to select 5 grant recipients among 100 applicants from multiple fields, and do this in limited time and in the presence of other commitments, the committee may consult bibliometric measures to inform their assessment of the scientists' work.

Since the use of citation-based metrics is likely to persist, we are motivated to design better metrics that would address the needs of the metrics' users (e.g., funding committees and fresh PhD students choosing an advisor) while aligning the incentives of scientists with the goals of the broader community.
To this end, we aim to address two drawbacks of commonly applied measures.
First, measures such as the $h$-index attach no cost to publication and thus incentivize opportunistic publishing: after all, publishing more can only improve the metrics, and even a ``publon'' can eventually accrue enough citations to increment the researcher's $h$-index. Such publication inflation can unnecessarily increase the number of papers that the research community must process (i.e., review, verify, and subsequently take into consideration) and divert the researcher's own attention from more substantial contributions that the researcher may be capable of.
Second, common metrics are designed to measure lifetime contribution rather than the impact of the scientist's recent work. A scientist who is long past their prime can have a higher $h$-index than a younger researcher who has revolutionized the field over the last few years. This renders such metrics inappropriate for answering questions that they are often called upon to inform, such as ``Who will convert our funding into the most impactful research over the next five years?'' and ``Who will guide my PhD dissertation into the most fruitful research avenues at this time?''

A number of citation-based measures do focus on taste rather than productivity.
We use `taste' to refer to a researcher's consistency in producing high-impact contributions, as quantified by citations. This is akin to `precision' in pattern recognition.
Examples of relevant measures are the mean number of citations across all of the researcher's papers~\citep{Lehmann2006} and the median number of citations for publications in the researcher's $h$-core~\citep{Bornmann2008}. However, these measures
can reward `one-hit wonders' and disincentivize exploration after a significant result. We seek a metric that will assign a higher value to a researcher who published 10 papers that are each cited 1,000 times than to a researcher who only published one such paper: the former demonstrated not only the ability to make a significant contribution (as measured by citations) but to do so repeatedly.
One possibility is to average the number of citations over the researcher's $k$ most highly cited papers. But this introduces a parameter $k$ that may need to be tuned to different values for different fields~\citep{Waltman2016}. A key characteristic responsible for the broad adoption of the $h$-index is the freedom from tuning such parameters.

In this report, we introduce a citation-based measure that rewards both taste and productivity. Let ${\sS = \{ s_i \}_{i=1}^P}$ be the set of publications produced by the researcher within a period of interest. Here $s_i$ is a publication in $\sS$ and the index $i$ iterates from $1$ to $P$, where $P$ is the total number of publications in the set. For publication $s_i$, let $C_i$ be the number of citations to it and let $A_i$ be its number of authors. The presented measure, CAP, is defined as follows:
\begin{equation}
  \mathrm{CAP} = \sum_{i=1}^P \big[ C_i - A_i - P \, \ge \, 0 \big] ,
  \label{eq:CAP}
\end{equation}
where $[\cdot]$ is the Iverson bracket. ($[x]$ is 1 if $x$ is true and 0 otherwise.)

To evaluate CAP for a researcher for a given year $Y$, we take the set of publications produced by the researcher in the 5-year period up to the year $Y-2$.
This 5-year period is set in accordance with findings that publications typically reach their peak citation rate within 5 years~\citep{Petersen2014}.
(\citet[\S5.5]{Waltman2016} reviews studies on citation windows and discusses settings ranging from one to five years, but see \citet{Ke2015} for an analysis of delayed recognition.)
$C_i$ is defined to be the number of citations accumulated by publication $s_i$ through the end of year $Y$. Thus CAP for year $Y$ cannot be computed until the following year, akin to the journal impact factor~\citep{Garfield2006}.
The 5-year period from which publications $s_i$ are drawn ends at year $Y-2$ rather than $Y$ to give publications a two-year grace period to begin accumulating citations.
Without the grace period, a publication would immediately increment the threshold $P$, without sufficient time to make an offsetting contribution to the count $C_i$. This would penalize bursts of productivity in the short term, even if the researcher produces consistently outstanding work.

The logic of CAP is that each publication increments the threshold $P$. This is the threshold that must be passed in order for the publication to contribute to the CAP. Furthermore, each publication increments the threshold $P$ for \emph{all} publications. Thus publication has cost. Each publication raises the bar that the researcher's work must meet. If the researcher publishes 10 papers within the five-year interval, CAP counts the number of papers
for which $C_i - A_i$ is at least 10. If the researcher publishes 100 papers, CAP counts papers for which the number of citations adjusted by the number of authors is at least 100.

The threshold $P$ acts as a soft incentive for the researcher to consider whether a publication is a useful addition to the literature. Each publication imposes a small externality on the research community, which must review and verify the work. Metrics that attach no cost to publication incentivize opportunistic production that does not take this externality into account. CAP models this externality in the threshold $P$. The metric is self-tuning and the threshold is set by the researchers themselves: the more they publish, the higher their CAP can be, but the bar that their publications must meet is also higher.

Unlike metrics such as the mean number of citations, CAP rewards productivity: a researcher who produces 50 outstanding papers in five years can have a CAP of 50, while the CAP of a researcher who produces 10 papers during this time will be at most 10. This reward for productivity is modulated by the researcher's taste: a researcher who produces 50 outstanding papers alongside hundreds of duds will do less well than their peer who was able to focus and devote their attention to their most impactful work.

CAP combines three quantities in a single formula: the number of citations $C_i$, the number of authors $A_i$, and the number of publications $P$. The number of authors $A_i$ is incorporated to mitigate the inflationary effects of hyperauthorship~\citep{King2012,Castelvecchi2015}. The adjustment by the number of authors in CAP is softer than in fractional allocation measures such as $h$-frac~\citep{KoltunHafner2021}: additive rather than multiplicative. This additive adjustment is more rewarding of collaboration: a publication coauthored by five researchers needs to accrue 5 additional citations to meet a given threshold, rather than $5 \times$. This small additive factor is unlikely to discourage innovative teams~\citep{Wuchty2007,Milojevic2014,Wu2019}.

CAP shares the key advantages of the $h$-index. It does not require tuning parameters, has a clear interpretation, and produces a single number that can be used for comparison. CAP also addresses the two drawbacks we have discussed. First, it rewards taste: the researcher's ability to focus on impactful work. The reward can be magnified by productivity, but productivity must act in conjunction with taste and can lower CAP if production outpaces the researcher's ability to identify opportunities for impactful contribution. The second key characteristic of CAP is its focus on recent work. CAP aims to quantify the researcher's taste and productivity in the recent past. While $h$-frac may be used to inform assessment of cumulative lifetime contribution~\citep{KoltunHafner2021}, CAP is designed to inform assessment of the researcher's present ability to effectively allocate resources.

\section{Results}

\begin{figure}[!ht]
  \centering
  \begin{subfigure}[t]{0.5\textwidth}
		\centering
    \caption{}
    \label{fig:dist-2020-overview}
    \includegraphics[width=\textwidth]{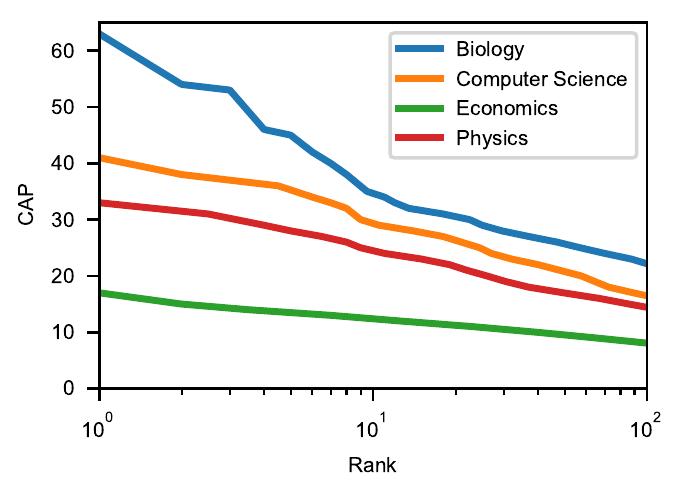}
	\end{subfigure}\\
  \begin{subfigure}[t]{\textwidth}
    \centering
    \caption{}
    \label{fig:dist-2020-tabs}
    \begin{minipage}{\textwidth}
      \begin{center}
      {\small
      \begin{tabularx}{0.45\textwidth}{rXl}
        \multicolumn{3}{c}{Biology} \\
        \toprule
        \# & Name & CAP\\
        \midrule
        1 & Eric Lander & 63 \\
        2 & Feng Zhang & 54 \\
        3 & Gad Getz & 53 \\
        4 & Michael S. Lawrence & 46 \\
        5 & Stacey Gabriel & 45 \\
        6 & Carl June & 42 \\
        7 & Peer Bork & 40 \\
        8 & Gordon Freeman & 38 \\
        9 & Simon I. Hay & 35 \\
        10 & Levi Garraway & 35 \\
        \bottomrule
      \end{tabularx}
      \hspace{6mm}
      \begin{tabularx}{0.45\textwidth}{rXl}
        \multicolumn{3}{c}{Computer Science} \\
        \toprule
        \# & Name & CAP\\
        \midrule
        1 & Ross Girshick & 41 \\
        2 & Xiaoou Tang & 38 \\
        3 & Jian Sun & 37 \\
        4 & Trevor Darrell & 36 \\
        5 & Xiaogang Wang & 36 \\
        6 & Jitendra Malik & 34 \\
        7 & Kaiming He & 33 \\
        8 & Cordelia Schmid & 32 \\
        9 & Abhinav Gupta & 30 \\
        10 & Chen Change Loy & 29 \\
        \bottomrule
      \end{tabularx}\\[5mm]
      \begin{tabularx}{0.45\textwidth}{rXl}
        \multicolumn{3}{c}{Economics} \\
        \toprule
        \# & Name & CAP\\
        \midrule
        1 & Michael Greenstone & 17 \\
        2 & David Roubaud & 15 \\
        3 & Raj Chetty & 14 \\
        4 & Nick Bloom & 14 \\
        5 & Derek Headey & 13 \\
        6 & John Van Reenen & 13 \\
        7 & Agnes Quisumbing & 13 \\
        8 & Sadok El Ghoul & 13 \\
        9 & Jarrad Harford & 13 \\
        10 & Daron Acemoglu & 12 \\
        \bottomrule
      \end{tabularx}
      \hspace{6mm}
      \begin{tabularx}{0.45\textwidth}{rXl}
        \multicolumn{3}{c}{Physics} \\
        \toprule
        \# & Name & CAP\\
        \midrule
        1 & Xiaodong Xu & 33 \\
        2 & Hongjie Dai & 31 \\
        3 & Riccardo Comin & 31 \\
        4 & Immanuel Bloch & 29 \\
        5 & Antonio H. Castro Neto & 28 \\
        6 & Hongming Weng & 27 \\
        7 & Wu Zhou & 27 \\
        8 & Ilya Belopolski & 26 \\
        9 & Nasser Alidoust & 25 \\
        10 & Pablo Jarillo-Herrero & 24 \\
        \bottomrule
      \end{tabularx}
      }
    \end{center}
    \end{minipage}
  \end{subfigure}
  \caption{CAP 2020 in four fields of research. (a) Distribution of CAP values for 100 researchers with the highest CAP in 2020 in each field. (b) Names of 10 researchers with the highest CAP in 2020 in each field. Researchers with the same CAP are ordered by $C$ (total number of citations within the same time frame).}
  \label{fig:dist-2020}
\end{figure}

Figure~\ref{fig:dist-2020-overview} shows the distribution of CAP values for 100 researchers with the highest CAP in each of four fields: biology, computer science, economics, and physics.
Here we report CAP for the year 2020. The top 100 authors have CAP values of 22 and higher in biology, 16 and higher in computer science, 14 and higher in physics, and 8 and higher in economics. Figure~\ref{fig:dist-2020-tabs} lists the 10 researchers with the highest CAP in each field in 2020.

\begin{figure}[!ht]
  \centering
  \begin{subfigure}[t]{0.5\textwidth}
    \centering
    \caption{}
    \label{fig:temporal-cross}
    \includegraphics[width=\textwidth]{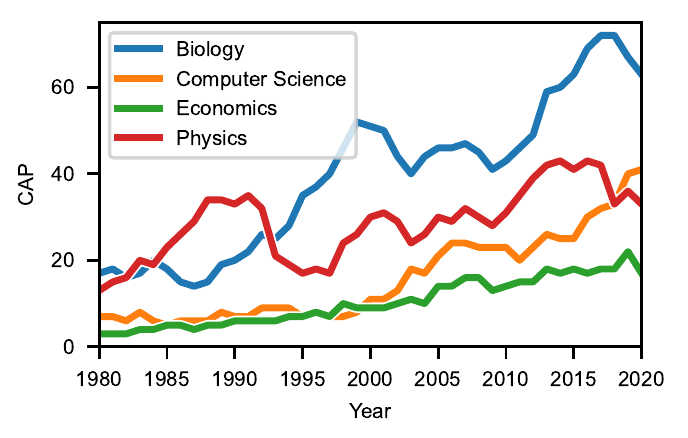}
  \end{subfigure}\\[-3mm]
  \begin{subfigure}[t]{0.535\textwidth}
    \centering
    \caption{}
    \label{fig:temporal-area}
    \includegraphics[height=15.5cm]{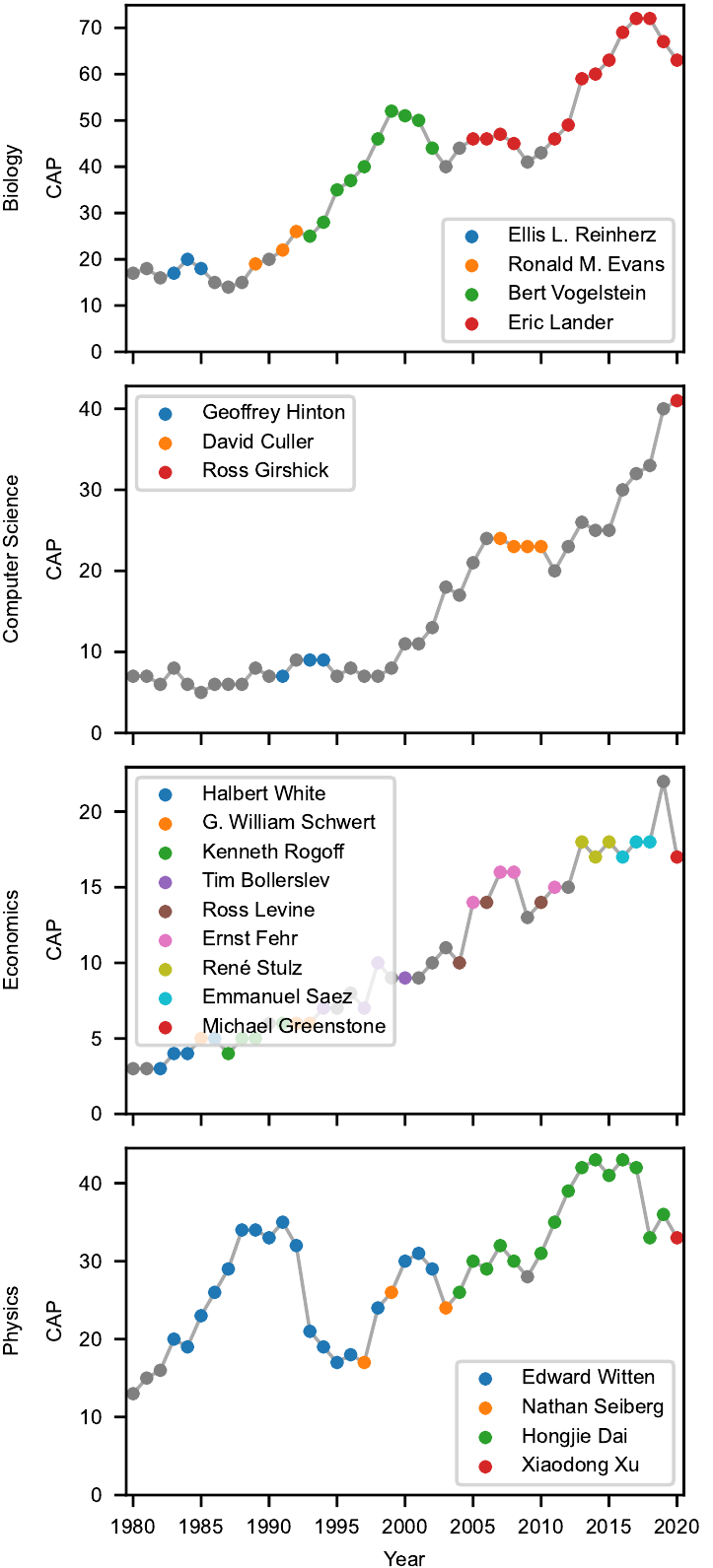}
  \end{subfigure}
  \begin{subfigure}[t]{0.457\textwidth}
    \centering
    \caption{}
    \label{fig:temporal-ind}
    \includegraphics[height=15.5cm]{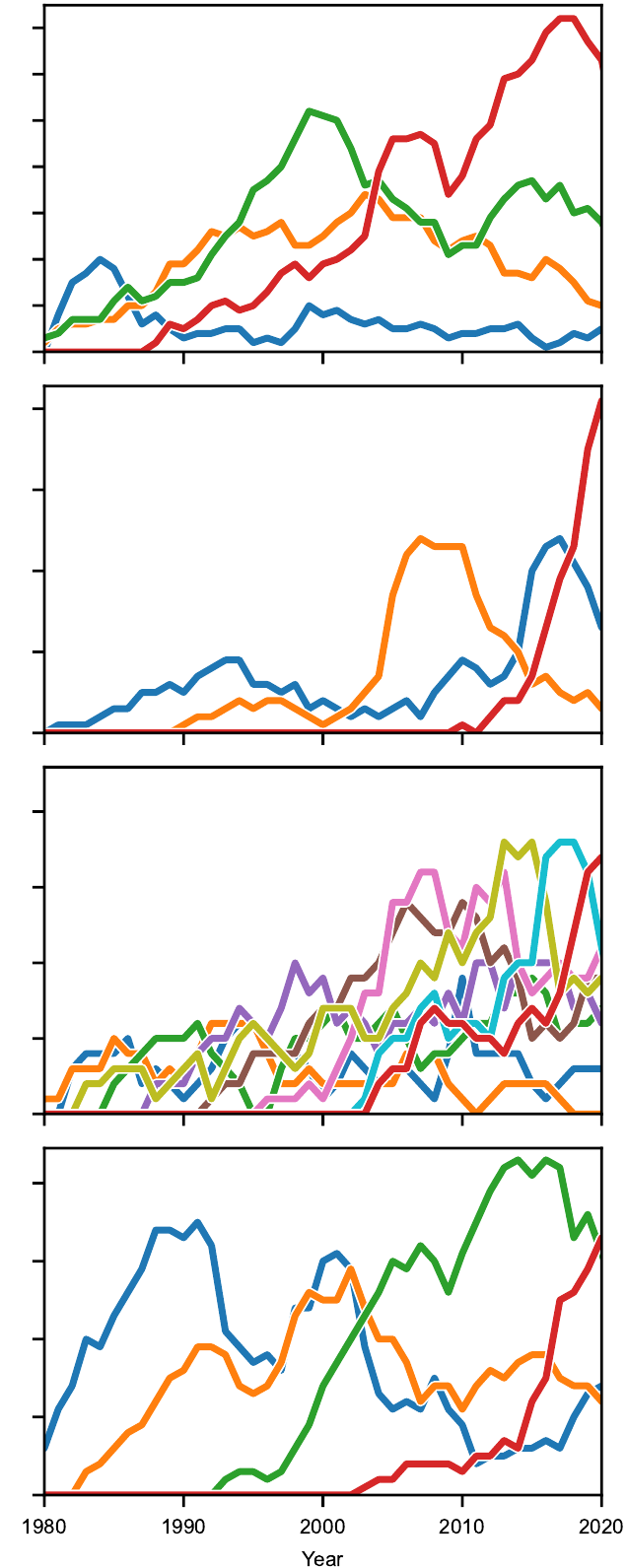}
  \end{subfigure}
  \caption{Temporal dynamics of CAP. (a) Highest CAP in each field over time. (b) Researchers in each field are highlighted if they have set the highest CAP in their field for at least 3 years or have the highest CAP in 2020. (c) Individual CAP trajectories for these researchers.}
  \label{fig:temporal}
\end{figure}

Figure~\ref{fig:temporal-cross} shows the evolution of the highest CAP in each field over the past 40 years. Here the year $Y$ is swept from 1980 to 2020. Figure~\ref{fig:temporal-area} breaks these curves down per year and lists a number of researchers who have set the highest CAP value in their field during this time. Figure~\ref{fig:temporal-ind} plots individual CAP trajectories for these researchers.

\begin{figure}[!ht]
  \centering
  \begin{subfigure}[t]{\textwidth}
    \centering
    \caption{}
    \label{fig:scatter-pubs}
    \includegraphics[width=\textwidth]{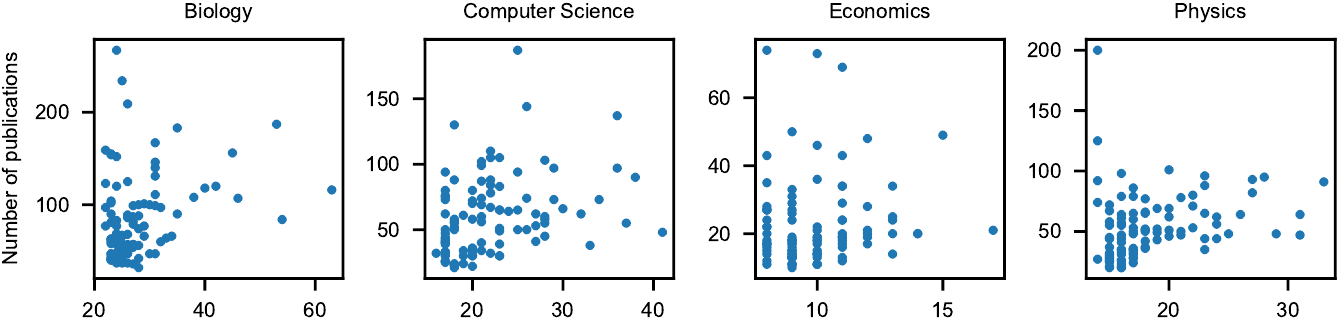}
  \end{subfigure}\\[-2mm]
  \begin{subfigure}[t]{\textwidth}
    \centering
    \caption{}
    \label{fig:scatter-authors}
    \includegraphics[width=\textwidth]{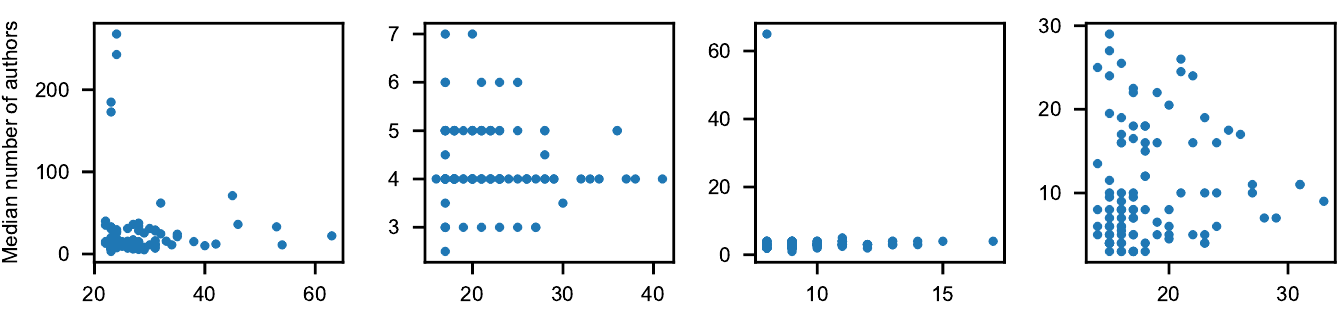}
  \end{subfigure}\\[-2mm]
  \begin{subfigure}[t]{\textwidth}
    \centering
    \caption{}
    \label{fig:scatter-age}
    \includegraphics[width=\textwidth]{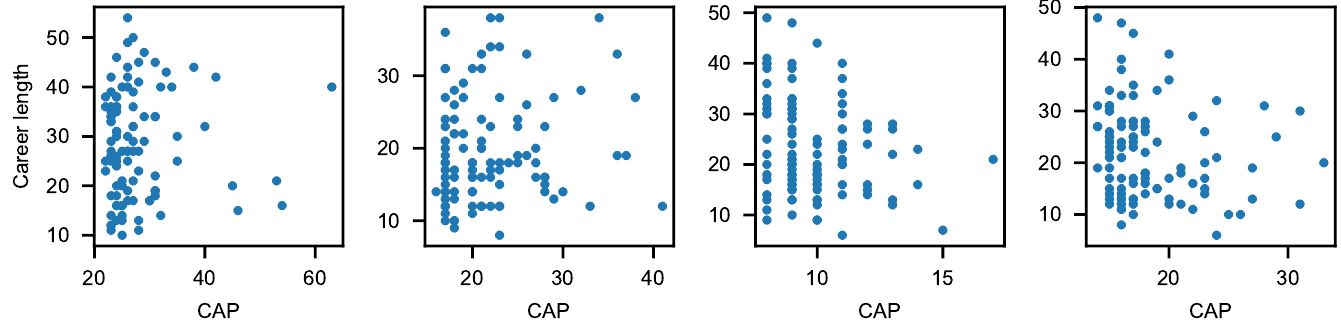}
  \end{subfigure}
  \caption{Scatter plots that relate CAP 2020 to (a) number of publications produced during 2014--2018, (b) median number of authors for these publications, and (c) career length, measured in years since first publication. Each scatter plot shows 100 researchers with the highest CAP in their field in 2020.}
  \label{fig:scatter}
\end{figure}

Figure~\ref{fig:scatter} provides scatter plots that relate CAP and a number of factors for 100 researchers with the highest CAP in 2020 in each field. Figure~\ref{fig:scatter-pubs} illustrates the relationship of CAP and the number of papers published during the corresponding 5-year period (2014--2018).
Figure~\ref{fig:scatter-authors} shows the relationship of CAP and the median number of authors for papers published during this period.
Figure~\ref{fig:scatter-age} examines the relationship of CAP and career length (measured in years since first publication).
CAP is largely uncorrelated with these three factors. The Pearson correlation of CAP with each of these factors over all fields is $-0.07 \le r \le 0.24$.

\begin{figure}[!ht]
  \centering
  \includegraphics[width=\textwidth]{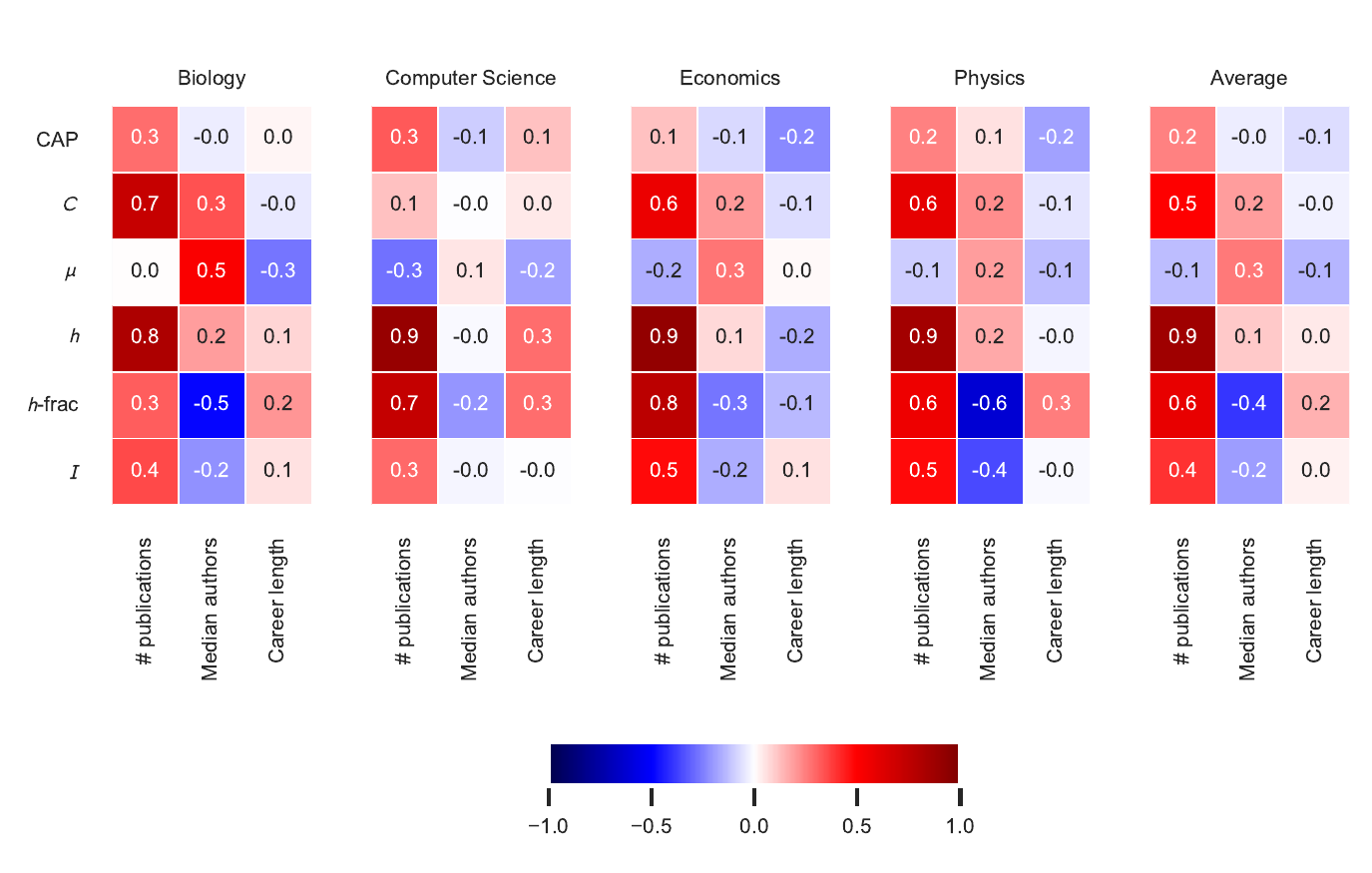}
  \caption{Correlation of citation-based measures with publication rate, median team size, and career length. The datasets and definitions are the same as in Figure~\ref{fig:scatter}. In addition to CAP, we evaluate $C$ (total number of citations), $\mu$ (average number of citations), $h$ (the $h$-index), $h$-frac, and the composite indicator of Ioannidis et al.~($I$). All measures are evaluated for publications produced during 2014--2018 and citations accrued through the end of 2020.}
  \label{fig:corr-PAL}
\end{figure}

Figure~\ref{fig:corr-PAL} examines these correlations quantitatively. Each column shows the correlation of a citation-based measure with publication rate, median number of authors per paper, and career length. This figure uses the same datasets and definitions as Figure~\ref{fig:scatter}. In addition to CAP, we examine a number of other citation-based measures, all computed on the same set of publications produced during 2014--2018.
Overall, the $h$-index is very highly correlated with publication rate ($r = 0.87$ over all fields). Thus, while the $h$-index was designed to provide a more refined measure of productivity than simply counting the number of publications, for high-performing researchers the two measures are statistically nearly indistinguishable. $h$-frac and $C$ (total number of citations) are also strongly correlated with publication rate ($r = 0.59$ and $0.51$ over all fields). CAP and $\mu$ (the average number of citations) are largely uncorrelated with publication rate (${r = 0.24}$ and $-0.13$ over all fields, respectively).

The correlations of citation-based measures and the median team size in Figure~\ref{fig:corr-PAL} are weaker. ($-0.40 \le r \le 0.26$ over all fields.) The average number of citations $\mu$ is most positively correlated with median team size (${r = 0.26}$), $h$-frac has the strongest negative correlation with median team size (${r = -0.40}$), and CAP is uncorrelated with median team size (${r = -0.04}$).

All measures in Figure~\ref{fig:corr-PAL} are largely uncorrelated with career length. ($-0.14 \le r \le 0.15$ over all fields.) This is consistent with findings that high-impact work can be produced with the same probability at any stage of a scientist's career~\citep{Sinatra2016}.

\begin{figure}[!ht]
  \centering
  \includegraphics[width=\textwidth]{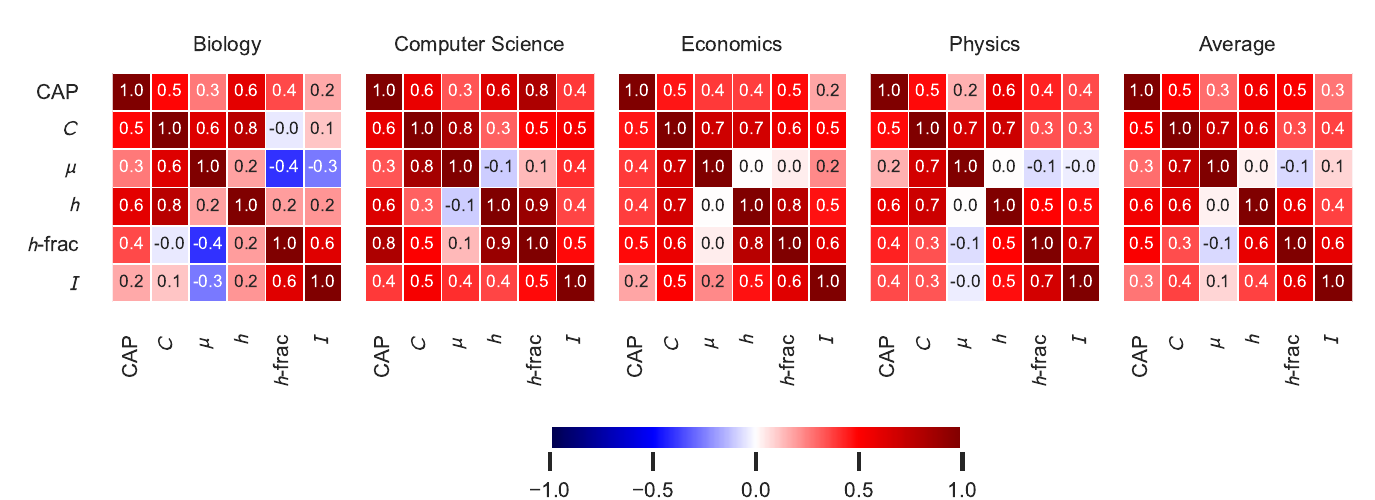}
  \caption{Correlations between citation-based measures: CAP, $C$ (total number of citations), $\mu$ (average number of citations), $h$ (the $h$-index), $h$-frac, and the composite indicator of Ioannidis et al.~($I$). The measures are computed as in Figure~\ref{fig:corr-PAL}.}
  \label{fig:corr-measures}
\end{figure}

Figure~\ref{fig:corr-measures} shows the correlation of CAP, $h$, $h$-frac, $\mu$ (average number of citations per paper), $C$ (total number of citations), and the composite indicator of~\citet{Ioannidis2016}, denoted by $I$. All measures are defined as in Figures~\ref{fig:scatter} and~\ref{fig:corr-PAL} and evaluated on the same datasets. The strongest correlations are between $C$ and $\mu$, $C$ and $h$, $I$ and $h$-frac, and $h$ and $h$-frac. ($r = 0.71$, $0.61$, $0.60$, and $0.59$ for these four pairs when averaged over all fields.)

\section{Discussion}
\label{sec:discussion}

\paragraph{Sensitivity to $P$.}
CAP takes more information into account than the $h$-index, and is thus sensitive to the accuracy of this information. Specifically, CAP is sensitive to the number of publications $P$, which determines the threshold that all publications must meet. The number of publications $P$ derived from publication lists in data platforms such as Scopus, Google Scholar, and Semantic Scholar is often wrong. Some platforms, such as Scopus, may not index all relevant publications, such as publications in conferences that are not tracked by the platform. This reduces the set of publications, the number of citations, and the threshold $P$, yielding overall lower metrics in Scopus datasets. An even bigger problem are spurious ``publications'' automatically added to researchers' profiles in the data platforms. These may be tables of contents from journal issues that are incorrectly classified as publications co-authored by all authors listed, prefaces to special issues edited by authors, supplemental materials to publications, or publications that were not authored by the researcher at all but were incorrectly associated with the researcher's profile. Such spurious publications are present to various extents in all data platforms, where they unnecessarily raise the threshold $P$. Google Scholar appears particularly susceptible to associating spurious entries with researchers' profiles. Researchers can edit their publication lists on Google Scholar, but some do not make use of this ability.
When a researcher's profile is cluttered with spurious entries and is not maintained by the researcher, their CAP may be significantly affected. (This sensitivity to the number of listed publications is shared by $\mu$, the mean number of citations.) All data in our report are derived from Scopus, which is more reliable in this regard.

\begin{figure}[!ht]
  \centering
  \includegraphics[width=\textwidth]{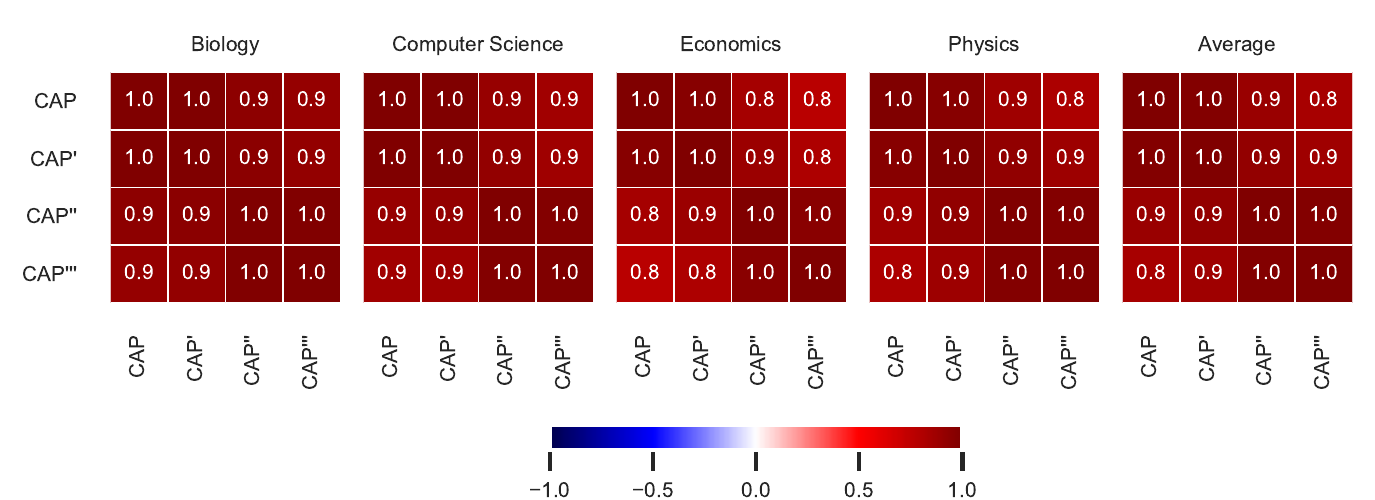}
  \caption{Sensitivity to noisy publication lists and values of $P$.  Correlation coefficients of a number of variants of CAP that assess sensitivity to noise (see text for definitions). The high correlation indicates a degree of stability in the computation of CAP.}
  \label{fig:sensitivity-to-p}
\end{figure}

We can assess the sensitivity to $P$ by estimating $P$ in a number of ways and measuring the stability of the computed CAP values. Since spurious ``publications'' automatically associated with authors' profiles often have very low citation numbers, we can try to prune them by pruning all entries with low citation counts. We test a number of variants: retaining all publications with non-zero citations (CAP'), retaining the most highly-cited publications that collectively make up $99\%$ of the total citation count (CAP''), and retaining the most highly-cited publications that collectively make up $98\%$ of the total citation count (CAP'''). Of course, genuine publications that have low citation counts may be pruned by these procedures as well. Figure~\ref{fig:sensitivity-to-p} shows the correlations between these CAP variants. The values are highly correlated, indicating a degree of stability to noisy publication lists.
We encourage data platforms to invest in automated data cleaning, and researchers to prune spurious entries from their profiles on platforms that provide access, such as Google Scholar.

\paragraph{Sensitivity to $A_i$.}
CAP is also sensitive to the accuracy of $A_i$, the number of authors of each publication $s_i$. (CAP shares this sensitivity with $h$-frac.) The set of authors is not always correctly indexed by all data platforms. In particular, Google Scholar appears to often cap the number of authors at 150, listing only up to 150 authors for a publication. Thus estimates of $A_i$ derived from Google Scholar can be wrong by an order of magnitude, for example for consortium publications with thousands of authors~\citep{King2012,Castelvecchi2015}. This is significant because hyperauthors and consortium publications are prominent among highly-cited authors in physics and biology~\citep{KoltunHafner2021}.
This is one of the key reasons for our use of Scopus rather than Scholar as the default data platform.
We encourage data platforms to correctly index the set of authors, even for hyperauthored publications.

Given that the incorporation of the number of authors $A_i$ in the definition of CAP induces sensitivity to this number's accuracy, would it not be better to omit this term altogether? A simpler metric, CP, can be defined that balances $C_i$ against $P$, without adjusting $C_i$ by $A_i$. Using the notation of Equation~(\ref{eq:CAP}),
\begin{equation}
  \mathrm{CP} = \sum_{i=1}^P \big[ C_i - P \, \ge \, 0 \big] .
  \label{eq:CP}
\end{equation}
CP is more compact than CAP and does not require authorship data. CAP and CP are strongly correlated in our datasets: $r = 0.76$, $0.97$, $0.82$, and $0.86$ for biology, computer science, economics, and physics, respectively, using the same data as Figure~\ref{fig:corr-measures}.
However, these values are computed for 100 researchers with the highest CAP in each field, and CAP already takes authorship into account through the term $A_i$.
If instead we consider 100 researchers with the highest CP in each field, the correlations between CAP and CP drop to 0.53, 0.93, 0.68, and -0.04, respectively.
The correlations drop the most in physics and biology, where hyperauthorship is common.
We use CAP rather than CP as the default measure in this study because of its resilience to hyperauthorship.

\begin{figure}[ht]
  \centering
  \includegraphics[width=0.5\textwidth]{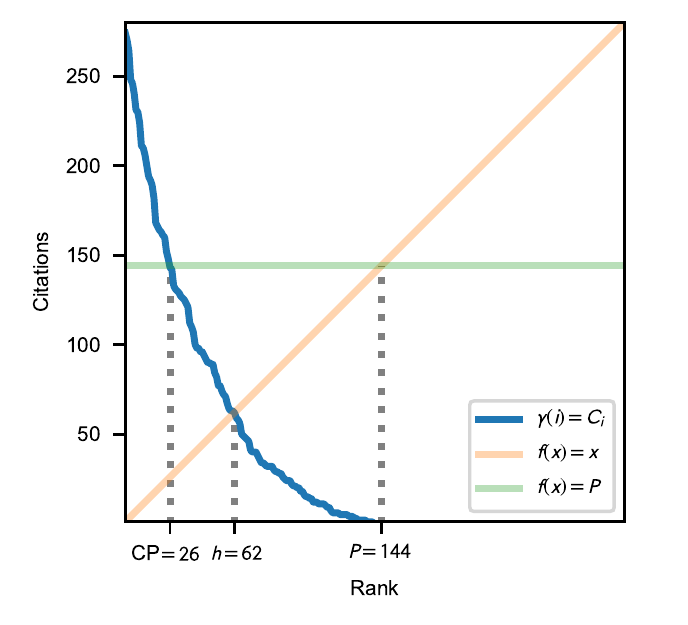}
  \caption{Example rank-citation profile with number of publications $P=144$. The $h$-index (${h=62}$) is given by the intersection of the rank-citation curve $\gamma(i)=C_i$ and the line $f(x)=x$. The intersection of the rank-citation curve and the line $f(x)=P$ yields $\mathrm{CP} = 26$.}
  \label{fig:rank-citation}
\end{figure}

\paragraph{Relationship to the $h$-index.}
The simplified measure CP can be used to draw an interesting relationship between CAP and the $h$-index. In the notation of Equation~(\ref{eq:CAP}), let the set $\sS$ of publications be ranked by the number of citations, such that $C_i \ge C_j$ when $i < j$. The function $\gamma(i) = C_i$ is known as the rank-citation profile~\citep{Petersen2011}. See Figure~\ref{fig:rank-citation} for an illustration. The $h$-index corresponds to the intersection of $\gamma$ with the function $f(x)=x$; that is, $\gamma(h)=h$. (Assume for simplicity of exposition that the function is continuous.) CP corresponds to the intersection of $\gamma$ with the function $f(x)=P$; that is, $\gamma(\mathrm{CP}) = P$. Since $h \le P$ and the function $\gamma$ is monotonically decreasing, we obtain the following relationship:
\begin{equation}
  \mathrm{CAP} \le \mathrm{CP} \le h .
  \label{eq:CAP-H}
\end{equation}

\paragraph{Data quality.}
As discussed earlier in this section, existing data platforms suffer from numerous data quality issues. Many initiatives seek to address this~\citep{Haak2012,Ammar2018,Wan2019,Lo2020,Hendricks2020,PeroniShotton2020,TeklesBornmann2020,Wang2020-MAG}. Nevertheless, we have not been able to identify a combination of data sources that yields researcher profiles with accurate publication lists, authorship information, and citation data~-- let alone open ones. We encourage allocating resources towards improving open-access data sources to increase their coverage and accuracy.

\paragraph{Limitations of citation-based measures.}
Our work shares the limitations of all citation-based measures of scientific impact. Citation counts are an imperfect proxy for impact and are affected by numerous biases~\citep{Garfield1979,DanielBornmann2008}. The importance of a work may not be reflected in its citation count for many years after publication~\citep{Ke2015}. Nevertheless, citations are a readily available, broadly applicable indicator that is amenable to large-scale computational analysis and are perhaps the most widely used quantitative proxy for the impact of scientific work~\citep{Wuchty2007,Radicchi2008,Wang2013,Uzzi2013,Petersen2014,Sinatra2016}.

\paragraph{Are we publishing too much?}
The answer to the question we began with is not obvious. The scientific community acts as a distributed system that produces and integrates information. The system has coped with exponential growth, indicating a degree of robustness~\citep{Dong2017,Fortunato2018}. While some argue that the pace and significance of innovation have declined~\citep{Gordon2000,GemanGeman2016}, the past decade has brought genome editing~\citep{DoudnaCharpentier2014}, deep learning~\citep{LeCun2015}, and the detection of gravitational waves~\citep{LIGO2016}.

\citet{GemanGeman2016} lament that ``many papers turn out to be early ``progress reports,'' quickly superseded''. But is this necessarily bad? Publication of ``progress reports'' is not a new phenomenon and the present time may not be unusual in this respect~\citep{Merton1957}. And contemporary research in large-scale optimization indicates that optimization processes driven by frequent, small, noisy signals may be more efficient and robust than processes driven by larger and more accurate steps~\citep{Bottou2018}.

Nevertheless, there are concerns that science is not self-correcting and the present reward system is suboptimal~\citep{Ioannidis2012,Ioannidis2014,Sarewitz2016NewAtlantis,SmaldinoMcElreath2016,EdwardsRoy2017,Lindner2018}. We hypothesize that adjusting the metrics that are used to evaluate researchers can increase the efficiency of the scientific community and promote scientific discovery. We hope that CAP and the analysis presented in this report can usefully inform the design and use of such metrics.

\section*{Acknowledgements}

We thank Alexei A. Efros, John P.A. Ioannidis, Jon Barron, Pushmeet Kohli, Raia Hadsell, Richard Newcombe, Vincent Vanhoucke, and Yann LeCun for providing feedback on an earlier version of this report. We summarize and respond to a number of questions and comments in Appendix~\ref{sec:response}.

\bibliography{main}
\bibliographystyle{report}

\appendix

\section*{\Large Appendix}

\section{Materials and Methods}
\label{sec:materials}

We closely follow the protocol of~\cite{KoltunHafner2021} to construct a dataset of highly-cited researchers in the following fields: biology, computer science, economics, and physics.
However, instead of targeting 1,000 researchers in each field, we start with 5,000 researchers per field retrieved from Google Scholar.
We match Google Scholar and Scopus profiles and clean the lists as described by~\cite{KoltunHafner2021}. This yields 4,262 researchers in biology, 3,011 in computer science, 2,218 in economics, and 3,792 in physics.
We collect all Scopus publications with citation data for all 13,283 authors and discard ancillary publication types such as editorials and commentaries (\cite{scopus}).
This yields a total of 3,042,438 publications that are collectively cited 227,149,547 times. The distribution of publications and citations among the research fields is as follows: biology accounts for 38\% of the publications and 54\% of the citations, computer science for 22\% and 14\%, economics for 5\% and 3\%, and physics for 36\% and 30\%.
Our dataset offers yearly temporal granularity from 1970 onwards. This enables detailed evaluation of the temporal evolution of citation-based metrics.

\section{Response to Feedback}
\label{sec:response}

We have shared an earlier version of this report with a number of researchers and have received helpful feedback. Our correspondents are acknowledged in the Acknowledgements section. Here we summarize and respond to a number of questions and comments.

\vspace{\baselineskip}
\noindent
\textbf{Q:} It seems that CAP can be quite low for many researchers and a CAP of 0 will not be uncommon. Is the measure still useful in this regime?

\noindent
\textbf{A:} Indeed, many researchers have CAP in the range 0--5. In our computer science dataset of $\sim$3K researchers, $\sim$1K have CAP $\ge 6$. The rest have CAP $\le 5$, and hundreds have CAP $=0$. Quantization is a significant factor in this regime and CAP does not make finer-grained distinctions within large groups of authors that have the same low CAP. One implication is that CAP may not provide a ``gradient signal'' in certain situations: an author whose CAP is 0 may change their behavior, but their CAP may remain 0 for some time. Our hope is that CAP nevertheless provides valuable information due to its simplicity and interpretability.

\vspace{\baselineskip}
\noindent
\textbf{Q:}
Have you tried computing a ``lifetime CAP'', an instantiation of Equation~(\ref{eq:CAP}) for all publications produced during the researcher's career?

\noindent
\textbf{A:}
We have. This does not appear to yield an informative measure. The first issue is that the indexing of older publications and citations in the data platforms is very poor. The second issue is that ``lifetimes'' of different researchers differ in ways that impact the ``lifetime CAP''. A researcher who published for 20 years during 1960--1980 is not comparable to one who published for 20 years during 1980--2000: the former's work has had 20 more years to accumulate citations that can ``counterweigh'' their number of publications $P$. And both of these researchers are not comparable to one who is still active, who may be at a disadvantage because they have recent publications that haven't had as much time to accumulate citations. The CAP measure as we have presented it does not suffer from these issues because the set of publications $\sS$ is drawn from the same recent 5-year period for all researchers.

\vspace{\baselineskip}
\noindent
\textbf{Q:}
The top of the CAP ranking seems to be heavily populated by people who work in ``fashionable'' areas. Some of them publish a lot and don't seem particularly restrained in this regard. Does CAP really accomplish what you intended?

\noindent
\textbf{A:}
The top of the CAP ranking is necessarily correlated with popularity of research areas, as expressed in citation counts. This relates to the limitations of all citation-based measures, as discussed in Section~\ref{sec:discussion}. Citations are a limited measurement instrument and are indeed correlated with ``fashion''. As for specific individuals, we observe that the top of the CAP ranking includes individuals with diverse publication styles. For example, among the top 10 authors in computer science (Figure~\ref{fig:dist-2020}), the number of publications produced during 2014--2018 ranges from 38 to 137.
Our impression is that researchers who have a high number of publications and a high CAP have a combination of taste and productivity that attracts such high citation volumes that they counterbalance the high threshold $P$ these researchers set for themselves.

\vspace{\baselineskip}
\noindent
\textbf{Q:}
In computer science specifically, the top 10 researchers listed in Figure~\ref{fig:dist-2020} are all in AI. In fact, they are all in computer vision. Isn't this strange?

\noindent
\textbf{A:}
To some extent this has to do with the popularity of different fields, as reflected in the numbers of publications and citations.
But the prominence of computer vision is exaggerated by uneven coverage in the Scopus data platform, which appears to index computer vision venues more thoroughly than machine learning, for example. Our supplementary website, \url{https://cap-measure.org/},
also provides computer science rankings based on Google Scholar data. In the ranking based on Google Scholar data, the top 10 authors in computer science are also all in AI, but only 3 out of the 10 are in computer vision.
Our impression is that the dominance of AI research in the CAP ranking reflects the popularity of AI and the volume of publications and citations associated with AI research. The prominence of computer vision in the ranking derived from Scopus data is spurious and is an artifact of uneven coverage in the Scopus database.

\vspace{\baselineskip}
\noindent
\textbf{Q:}
Is CAP more inclusive than prior metrics? How does CAP reflect on younger researchers? What are the implications for under-represented groups, such as female researchers in computer science?

\noindent
\textbf{A:}
CAP appears to be particularly conducive to the advancement of young researchers. For example, among 100 authors with the highest CAP in computer science, 5 have career lengths $\le$10 years and 31 have career lengths $\le$15 years. (Career length is measured in years since first publication.) In contrast, when authors in the same field are ranked by their $h$-index, not a single researcher with career length $\le$15 appears in the top 100.

How much of this difference is due to the focus of CAP on recent work versus the lifetime accumulation of the $h$-index? To probe this we can consider an application of the $h$-index to the same set of publications as CAP (as in Figures~\ref{fig:corr-PAL} and~\ref{fig:corr-measures}). When evaluated on the same set of publications as CAP, the $h$-index ranks 0 computer science researchers with career length $\le$10 in the top 100, and 14 researchers with career lengths $\le$15. This is still much lower than 5 and 31, the corresponding numbers for CAP.

\begin{figure}[ht]
  \centering
  \includegraphics[width=0.75\textwidth]{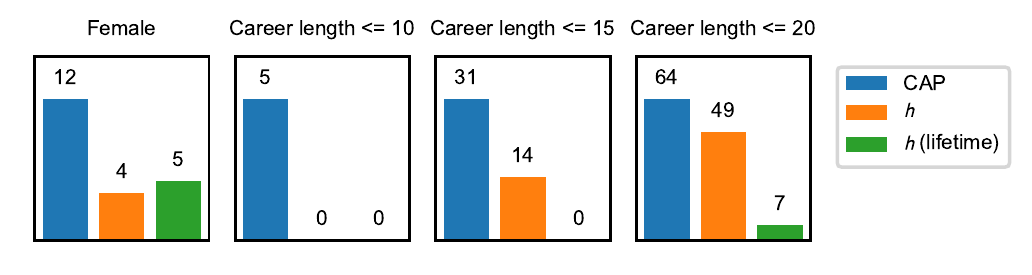}
  \caption{A preliminary examination of implications for young researchers and under-represented groups. Representation among the top 100 authors in computer science, ranked by different metrics. The metrics are CAP, $h$-index evaluated on the same set of publications as CAP, and the standard (lifetime) $h$-index.}
  \label{fig:inclusion}
\end{figure}

As a preliminary examination of implications for under-represented groups, we have manually annotated the data to quantify the inclusion of female computer science researchers. When authors are ranked by the $h$-index, 5 female researchers appear in the top 100 in computer science. This poor representation can also be observed in public rankings such as \url{https://www.guide2research.com/scientists/}. Evaluating the $h$-index on the same set of publications as CAP does not increase the representation: the corresponding ranking includes only 4 female researchers in the top 100. In contrast, a ranking by CAP includes 12 female researchers in the top 100 in computer science. This is summarized in Figure~\ref{fig:inclusion}. The data is consistent with the hypothesis that CAP can support the advancement of young researchers and historically under-represented groups.

\vspace{\baselineskip}
\noindent
\textbf{Q:}
Could CAP disincentivize risk taking?

\noindent
\textbf{A:}
CAP aims to incentivize researchers to think about the contribution made by each publication, and whether this contribution offsets possible negative externalities.
CAP appears to be quite compatible with thoughtful exploration, in part because low-impact publications are no longer taken into account after $\sim$7 years.

\vspace{\baselineskip}
\noindent
\textbf{Q:}
Could CAP incentivize some researchers to ``hide'' publications that are not well-cited, to try to make such publications ``disappear'' somehow?

\noindent
\textbf{A:}
We hope not. While this may be possible on data platforms such as Google Scholar, it is not possible on platforms such as Scopus. And fudging data can lead to strong reputational damage that outweighs whatever benefits may accompany a somewhat higher CAP.

\vspace{\baselineskip}
\noindent
\textbf{Q:}
In general, can broad adoption of CAP have unexpected consequences? Side-effects that you have not anticipated? Would it be possible to study this, perhaps simulating what happens if scientists adopt a measure such as CAP as an objective?

\noindent
\textbf{A:}
Yes. Broad adoption of measures and incentives does often have side-effects. We can observe the side-effects of existing measures, such as the $h$-index. Simulating possible effects of alternative measures is a fascinating research direction. We would like to see more analyses of the scientific community as a dynamical system that is shaped by incentives. Such analyses can use computational modeling (simulation) and mathematical techniques. \citet{SmaldinoMcElreath2016} is one stimulating work in this vein.

\vspace{\baselineskip}
\noindent
\textbf{Q:}
Do you really expect a new citation-based measure to shift the culture away from maximizing the production of publications?

\noindent
\textbf{A:}
Stranger things have happened.

\end{document}